\documentclass[12pt]{iopart}

\usepackage{graphicx}
\usepackage{dcolumn}
\usepackage{bm}
\usepackage{caption}
\usepackage{subfloat}
\usepackage{subfig}
\usepackage{float}
\usepackage{color}
\usepackage{setspace}
\usepackage{amssymb}
\usepackage{color,soul}
\usepackage{braket}

\begin{document}

 \title{The Concept of Time: A Grand Unified Reaction Platform}

\author{Hamidreza Simchi}
\ead{simchi@alumni.iust.ac.ir}
\address{Department of Physics, Iran University of Science and Technology, Narmak, Tehran 16844, Iran}
\address{Semiconductor Technology Center, P.O. Box 19575-199, Tehran, Iran} 

\date{\today}

\begin{abstract}
 The universe is things which change and called events. The events are matter and field. A boundary divides a system to things and environment. The things belong to the environment have no significant effect on the things belong to the system. The physical observables are the variations of things and it is always assumed that the conscious thing is placed in environment because the science cannot explain consciousness. There is not only an obligated minimum boundary between things (space) but also between past and future (present). The gravitational field has significant effect on these obligated minimums, especially at Planck scale. By using the above concept we introduce a grand unified reaction platform for categorizing the current physical paradigms and possible future explanation of the universe as a whole.  
\end{abstract}
\vspace{2pc}


\maketitle

\section{Introduction}

         From many years ago, people have been familiar with the concepts of matter and field. Einstein have explored the relativistic mechanics and shown that matter can convert to energy and vice versa. By exploring the quantum field theory, people had attributed specific quanta to each field. During the last century, people had tried to unify the existed different physical theories. Standard model unifies the matter and all known different fields, except gravitational field.  Some scientists believe that abortiveness in adding the gravitational field to standard model is referred to the concept of time [1, 2]. Nowadays, there are two different answers to the question about the reality of time.  We chose some important statements of each answer and rewritten them in table 1. 

{\begin{table}[ht]
\caption{Two different answers to the question about the reality of time} 
\centering 
\begin{tabular}{c c} 
\hline\hline 
Professor Barbour{[}1{]} & Professor Smolin{[}2{]} \\ [1ex] 
\hline 
The change of things is time. & Time is the most real   \\ 
 Time is simply a complex of rules & aspect of our perception of the world. \\
that govern the change. & Non-causal childrenTime is the most real aspect of our \\
Time is inferred from things. &perception of the world. Space is emergent and approximate. \\
Time is in the instant. & Laws of nature evolve in time .\\
\hline 
\end{tabular}
\label{table:nonlin} 
\end{table}

 \noindent However, Marchesini have evaluated the Barbour's and Smolin's answers through the lens of Henri Bergson's metaphysics of time [3]. He has explained how both ideas encounter with some paradoxes and in consequence run into dead ends [3]. The some chosen statements of Bergson's idea are rewritten in table 2. {[}3, 4, and 5{]}:

{\begin{table}[ht]
\caption{Marchesini's evaluation of Barbour's and Smolin's answers {[}3, 4, and 5{]}} 
\centering 
\begin{tabular}{c} 
\hline\hline 
Marchesini's evaluation through the lens of Henri Bergson's metaphysics of time {[}3{]} \\ [1ex] 
\hline 
Things precede nothingness. Creative order precedes disorder. \\ 
 Movement and change precede inertia and immobility. \\
Time must not be confused with space; to pass from one to the other one had only to change  \\
a single word: juxtaposition was replaced by succession. \\
As time passes, existence is merely added to what was already possible. \\
Our states of consciousness are continuous, indivisible, and interpenetrate one another.\\
The whole of the universe moves and changes much like our conscious.\\
\hline 
\end{tabular}
\label{table:nonlin} 
\end{table}

\noindent      In this paper, we intend to show how to define certain boundaries between physics in different dimensions, through which a conceptual shift paradigm occurs. Following this explanation, we intend to introduce a reference platform based on which the grand unified theory may be described and explained. Therefore, we review some current different concepts of time, Firstly. Then, we explain how a conceptual shift paradigm is happen when we consider the universe as a whole instead of a part of the universe. Also, we explain how some specific concepts, such as the level of awareness (LoA), obligated minimum boundary between things and obligated minimum boundary between past and future can help us in planning a grand unified reaction platform (GURP) for explaining the different concepts of time and thinking about the different physical paradigms.  The structure of article as follows: in section II, the different current concepts of time are reviewed. The concept of the grand unified reaction platform is explained in section III. The summary is provided in section IV.

\section{Different concepts of time}

\noindent   There are different explanations about the concept of time. In below, we choose and review the most important ones.

\subsection{Classical concept of time}

\noindent      If we assume that the world contains things and time ($t$) flows in the world then we can conclude that things change in time, have volume and occupy the space ($\overrightarrow{x}$). Therefore, time and space are different concepts. Since, the time is considered as a flux, we should divide the time axis to three different sections, called past, present (now), and future. Now, if $\ \mathrm{\Psi }\mathrm{(}\overrightarrow{x}\mathrm{)}$  stands for the state of a thing; its variation in time (its evolution) i.e.,$\mathrm{\ }\mathrm{\Delta }\mathrm{\Psi }\mathrm{(}\overrightarrow{x}\mathrm{)/}\mathrm{\Delta }t$  will provide us the future state of the thing. If we know the past of a thing (called boundary conditions or history in general) we will able to find its future. It means that we receive the information from the past of things and in consequence we called $\mathrm{\Psi }\mathrm{(}\overrightarrow{x}\mathrm{)}$  the retarded state of things. It means that we receive the information with some delays. Therefore, the world is like a big mechanical machine which works deterministically, based on the classical laws. It should be noted that if one knows the future state ($\mathrm{\Psi }\mathrm{(}\overrightarrow{x}\mathrm{))}$; he/she will able to find its past by reversing the axis of time and solving the deterministic classical equations (of course, if dissipative terms do not exist). It means that the future information is built by the past information. In this case, we called $\mathrm{\Psi }\mathrm{(}\overrightarrow{x}\mathrm{)}$  the advanced state of things. It means that the future information comes from the past. In the point of view, the evolution of things is absolute and is not relativistic. 

\noindent 

\noindent      But, we can rewrite the above scenario in a new style. We can assume that the world contains things and they move along their world line,$\ \gamma $, in the world. The world lines are made by the existence of the thing, their interaction with gravitational field and their relative evolution. They fill the entire world. Therefore the state of thing ($\mathrm{\Psi }\mathrm{(}\overrightarrow{x}\mathrm{,t))}$ is confined to change on its world line. Here, the time ($t$) has no physical meaning and can be changed, freely but things occupy the spacetime ($\overrightarrow{x},t$).  For calculating the time one should study the state of a clock (thing) when it moves along its world line. In the point of  view, the evolution of things is relativistic and is not absolute. It means that there is no difference between present, past, and future [2]. It should be noted that since the time ($t$) can be changed freely, one is able to disappear it from the classical Hamiltonian -- Jacobi formulation and find a time independent equation such as Wheeler- Dewitt (WdW) equation. Simply speaking, the WdW equation says $\hat{H}$${\Psi}=0$, where $\hat{H}$ is the Hamiltonian constraint in quantized general relativity and ${\Psi}$  stands for the wave function of the universe. Unlike ordinary quantum field theory or quantum mechanics, the Hamiltonian is a first class constraint on physical states. We also have an independent constraint for each point in space. Therefore, WdW equation does not describe a frozen world [6]. 

\noindent         Finally it can be concluded that, all other observable quantities change in space by passing the time or change in spacetime. Basically in the above both point of views, the world contains things and they change absolutely or relative to each other. It means that the other physical concepts are defined (confined) by the concept of time and space.  

\noindent

\subsection{Quantum concept of time}

\noindent         Although, the Schr\"{o}dinger equation is a quantum mechanical equation but it evolves in classical space when time is considered as a flux. Dirac and Kelin-Gordon equations have been found by using the relativistic  laws but both evolve in spacetime. On the right-hand side of the Heisenberg uncertainty relation between space coordinate and momentum of a particle, the time parameter ($t$) does not appear. It means that the priority is not important. Therefore, we encounter the classical concept of time or spacetime (called background dependent). By considering the Schrodinger, Dirac, Klein-Gordon,Yang-Mills and so on as a field function ($\mathrm{\Psi }\mathrm{(}\overrightarrow{x}\mathrm{,t))}$ and quantizing them by using the second quantization methods, the term $\delta (t-t^\prime)$ appears on the right-hand side of the Heisenberg uncertainty relation [7]. Therefore, the priority is important. But similar to the previous case, we encounter the background dependent case, again. 

\noindent        Also, we know that there is an uncertainty relation between energy and time in quantum mechanics which is $\Delta{E}\Delta{t}\geq\hbar/2$ and $(\Delta{E(t))^2}$$=<E(t)^2>-<E(t)>^2$. It means that the change in time is connected to the change in energy [2]. For understanding the relation let us consider an experiment. An electron, which is in its ground state, is excited to a higher energy level and after passing sometimes it comes back to the ground state. It can be shown that the staying time in the excited state ($\Delta t$) is proportional to the inverse of the energy difference between the excited and ground states ($\Delta E$) [8]. But, we are in the classical atmosphere of time else. Also, we can consider a particle with mass $M$ and energy$\ \ E=Mc^2$, where $c$ is the velocity of light. We can localize the particle in a sphere with radius$\ R\sim MG/c^2$ where$\ G$ is the gravitational constant. Using the Heisenberg uncertainty principle, it can be shown that it is not possible to localize anything with a precision better than the Planck length which is equal to ${10}^{-33}$\textit{cm} [9, 10]. It means that anything smaller than the Planck length is hidden inside its on mini-black hole [10]. This is called the length (Planck) scale. By using the Planck length, the minimum time can be calculated and called the Planck time which is equal to ${10}^{-44}$ seconds. These are the granules of quantum gravity [11]. Since, the quantum spacetime is a physical object and fluctuates; it can be in a superposition of different configurations (spacetime) [11]. Of course, due to the relational aspect of quantum physical variables, the quantum gravitational field does not have determined values until it interacts with something else [11].By ignoring the microscopic details of the world, a blurring is seen i.e., it is produced by the intrinsic quantum indeterminacy of things [11]. The time of physics is the expression of our ignorance of the world i.e., time is ignorance [11].

\noindent      Based on the quantum methodology, for finding the quants of space and area we should define the volume and area operators [9, 10]. It can be shown that there is a kinematical Hilbert space ${\mathcal{K}}_{Diff}$ which admits a basis of states in which certain area and volume operators are diagonal. It means that there is a spin network state which describes a quantized three-geometry [9, 10].It should be noted that a spin network state is not in space but it is space [9, 10]. Therefore, it introduces a background independent physics and in consequence we are in quantum atmosphere of space. It is well known that, the quantum dynamics of a particle is entirely described by the transition probability amplitudes $W\left(x,t,x^\prime,t^\prime\right)=\left\langle x\mathrel{\left|\vphantom{x e^{-\frac{i}{\hbar }H_0(t-t^\prime)} x^\prime}\right.\kern-\nulldelimiterspace}e^{-\frac{i}{\hbar}H_0(t-t^\prime)}\mathrel{\left|\vphantom{x e^{-\frac{i}{\hbar}H_0(t-t^\prime)} x^\prime}\right.\kern-\nulldelimiterspace}x^\prime\right\rangle =\left\langle x,t\mathrel{\left|\vphantom{x,t x^\prime,t^\prime}\right.\kern-\nulldelimiterspace}x^\prime,t^\prime\right\rangle $ which depend on two events $(x,t)$ and $(x^\prime,t^\prime)$ that bound a finite portion of a classical trajectory [9]. It should be noted that the argument of $W$ is the eigenstate of the corresponding Hamiltonian operator, $H_0$. Similarly, we can consider a spin network as the argument of $\ W$ which represents the possible outcome of a measurement of the gravitational field (or the geometry) on a closed three dimensional surface [9]. By defining a suitable scalar product, it can be shown that the transition amplitude between two states is simply their scalar product [9]. Here, H acts on the node of spin network and in consequence spinfoam is constructed [9, 10]. The transition amplitudes $W\left(s,s^\prime\right)\ $do not depend explicitly on time and in consequence introduces a new concept called the physic without time. It means that the theory allows us to calculate the relation between observables which is what we see and does not give us their evolution in terms of an unobservable quantity which is called time in classical physics. In the other words, there are no good clocks at the Planck scale [9]. The world is a network of events and is not things [11]. In classical physics time exists with many determinations and here the main concept is: things happen [11].The difference between things and events is that things persist in time but events have a limited duration [11]. 

\noindent       Finally it can be concluded that, the spin network is space, and its evolution by acting the Hamiltonian, constructs the spinfoam. But, it is not clear that how other observable quantities change in the spin network. For example, how can we study the quantum transport of a particle in a spin network and find its spinfoam? What may be the general laws and rules? Whether, the above explained theory can be only used for studying the some specific problems of quantum gravity and the structure of spacetime at Planck scale [12]? It means that we do not know how the other physical concepts are defined (confined) by the concept of the spin network. 

\noindent 

\subsection{Biological concept of time}

\noindent Three types of time can be defined in biology as follow:

\begin{enumerate}
\item  The number of complete cycles per unit time such as heart rate and metabolic rate

\item  Aging due to the cell dividing(splicing) process

\item  Internal  clock of body
\end{enumerate}

\noindent Many Biological variables can be defined as the product of three power function as below
\begin{equation} \label{GrindEQ__1_} 
\left[Y\right]=[M^{\alpha }\ .\ L^{\beta }\ .\ T^{\gamma }] 
\end{equation} 
where, $M$, $L$, and $T$ stand for mass, length, and time, respectively [13]. For comparing the empirical findings with calculated values, the biological variable $Y$ can be expressed as 
\begin{equation} \label{GrindEQ__2_} 
Y={\mathrm{log} a+b\ .{\mathrm{log} M\ }\ \ } 
\end{equation} 
which is called Huxley's allometric equation [13, 14]. Using the dimensional analysis, one can investigate the influence of earth's gravity on heart rate and metabolic rate. It can be shown that there are two solutions for allometric coefficient $b$  as below [13, 15]
\begin{equation} \label{GrindEQ__3_} 
b_M=\alpha +\frac{1}{3}\beta +\frac{1}{4}\gamma  
\end{equation} 
and
\begin{equation} \label{GrindEQ__4_} 
b_W=\frac{5}{6}\alpha +\frac{1}{3}\beta +\frac{1}{4}\gamma \  
\end{equation} 
where, $b_M$ and $b_W\ are\ $body mass and body weight coefficients, respectively [13, 15].      As an example, for metabolic rate $b_M=0.91$ and $b_W=0.75$ the difference  $\Delta b=0.17$ which is referred to the influence of the gravity [13]. It should be noted that the predominance of the cyclic nature of almost all functions leads to the granulation of time. It appears as flicker-fusion-frequency in neurophysiological realm, heart rate, respiratory rate, and specific metabolic rate in organ physiology [13]. 

\noindent        One of the strange natural phenomena is frozen wood frogs which have been seen in Alaska [16]: 

\noindent \textit{``The hearts of the frogs stop beating and their blood no longer flows. The freezing patterns help the frogs convert more of the glycogen stored in their liver into glucose. It is the high levels of glucose in the cell of frogs that keep them alive through the long, cold winter. The main function of the glucose is to keep water inside the cell. By making the cells super sweet with glucose, the frogs keep the water from leaving their cells. It should be noted that, they do not freeze totally solid, but they do freeze mostly solid. Two-thirds of their body water turns to ice.''}

\noindent \textit{ }Therefore, under very special and critical thermo dynamical conditions, the cyclical character of the biological time can be changed such as the heart beating of wood frogs in Alaska and in consequence can be set in laboratory.\textit{}

\noindent       There are two different ideas about whether ageing is a disease or not [17]. Aging can be considered as a natural process and consequently it is not a disease [17, 18, 19]. They believe that aging constitute a natural and universal process, while diseases are seen as deviation from normal state. Some other scientists believe that aging has specific causes, each of which can be reduced to a cellular and molecular level, and has recognizable signs and symptoms [17, 20]. However it is well known that, certain genes and pathways that regulate splicing factors play a key role in the aging process [21]:

\noindent  \textit{``The ERK and AKT pathways are repeatedly activated throughout life, through aspects of aging including stockticker DNA damage and the chronic inflammation of aging. By using specific inhibitors which are already used as cancer drugs in clinics, it is possible the activity of the ERK and stockticker ATK pathways are stopped and in consequence an increment in splicing factor is seen. It means that a better communication happen between protein and genes. It can cause a reduction in the number of senescent cells and reverse many of their features which have been linked to the aging process and in consequence leading to a rejuvenation of cells.'' }

\noindent     Some scientists plan to rejuvenate dogs using gene therapy. They plan to make animals younger by adding new stockticker DNA instructions to their bodies [22]. Also, the eroding effects of aging can be controlled by making a complex and protective shield which is made by combination of stem cell with anti-aging gene [23]. Therefore, the ageing character of the biological time can be changed and set in laboratory.  

\noindent     Another biological concept of time is referred to the body clock [24]:

\noindent \textit{``People belong to each time zone of the earth have a specific rhythms which is in sync with the day-night cycle of the zone.  The rhythms dictate times for eating, sleeping, hormone regulation, body temperature variations and other functions. The rhythms have been changed by rapid long-distance trans-meridian (east-west or west-east) travel. The adjustment speed of jet lagged body depends on the individual as well as the direction of travel. It has been shown that the jet lag is a chronobiological problem.'' }

\noindent Therefore, it can be concluded that the body clock is some kind of complete cycles per day which is adjusted by some specific duration and depends on individual and travel direction. 

\subsection{ Cosmological concept of time}

\noindent      If the universe (world) is the variable things, a thing divides the world to two main parts which are called system (thing) and environment (here, background). Of course one can consider a set of things as a system and calls the remained part of the universe as its environment. Therefore, there is a boundary between system and its environment. By adding more things to the system, the system inflates and its environment condensates (respect to its previous condition). The boundary can be defined as: ``\textit{things which are placed outside the boundary have negligible effect on the system and the effect of things which are placed inside the boundary should be considered when we want to study the specific characteristics of the system. Outside a boundary contains enviroment i.e., other things''.} For example, a confined gas in a region (box) can be considered as a system. By attaching a thermometer on the outside boundary of the system (walls of the box), one can measure the temperature (kinetic energy of gas molecules) of the box. By setting a clock outside the box, it is possible to study the time variation of the temperature when the boundary of system (the volume of the box) increases or decreases. Also we can calculate the entropy of the system. Conlon has defined the entropy of an object as a measure of its number of degrees of freedom i.e., ``the total number of ways to rearrange its internals while keeping its external unaltered''. For example, the entropy of gas is a measure of the total number of way the gas molecules can arrange themselves within box'' [12]. It means that by increasing the number of rearrangement the entropy increases and in consequence the system is more stable than before. Smolin has defined the entropy as: ``how many microstates could give the same macrostate. The entropy of a building is the measure of the number of different ways to put the parts together to realize the drawing of the architect''[2]. Of course, the entropy is proportional to information, inversely. For example, one can use four small equilateral triangles (case No.1) or three small isosceles triangles (case No.2) and make a bigger equilateral triangle by using them. Also, it is possible one make an equilateral triangle by using three triangles and one parallelogram (case No.3) (Fig.1). It is obvious that for manufacturing the case No.3 we need more information respect to the case No.2 and for manufacturing case No.1 we need less information respect to the case No.2. Also, the case No.3 is more orderless than case No.2 and the case No.1 is more order than case No.2. However, we can set the small triangles of Fig.1 (a) in many different ways and find the big triangle. But the number of setting the small triangles of Fig.1 (b) for finding the big triangle is less than Fig.1 (a) (i.e., we need more information). Similarly, the number of setting the small triangles of Fig.1 (b) for finding the big triangle is more than Fig.1 (c) (i.e., we need less information). Therefore, the information is proportional to the orderless (entropy), inversely. But it should be noted that in both Conlon's and Smolin's point of view, there is a boundary which separates the system from its environment.

\begin{figure*}[t!]
\includegraphics[width=\textwidth]{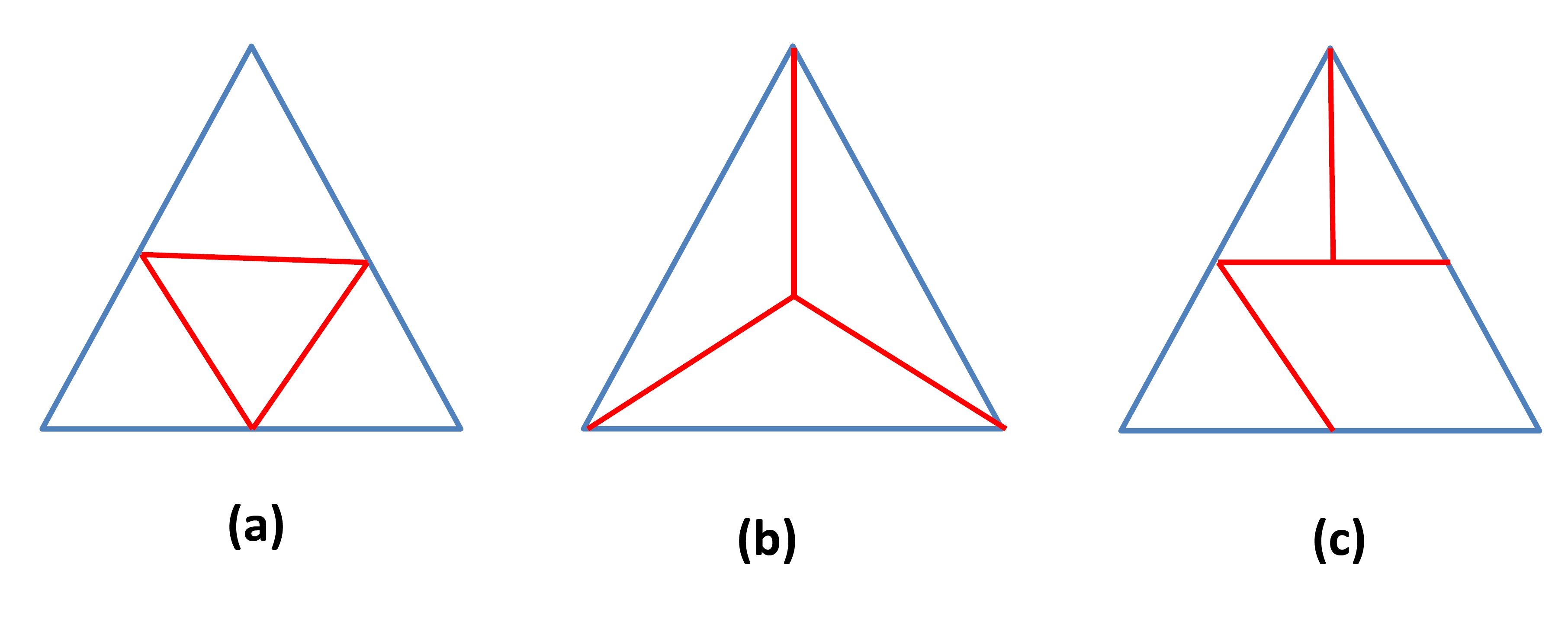}
\textbf{\caption{(Color online) A equilateral triangle composed by (a) four equilateral triangles, (b) three isosceles triangles, and (c) three triangles and one parallelogram}}
\label{fig:Fig1}
\end{figure*}

\noindent 

\noindent      The biggest problem about the universe as whole is: how can we define the boundary of the universe? If there is not an environment (the existence of other things which have not serious effects on the universe) how can we define a clock outside the universe and measure the time? How can we put a thermometer outside the universe and measure its temperature? How can we define the order or orderless, entropy and information when we are not able to define the environment (outside) of the universe? How can we define the inflation or condensation? It should be noted that, we cannot put a clock outside the universe and in consequence its time is zero and the energy of the universe is equal to zero too, based on the Heisenberg uncertainty principle. Therefore, it can be concluded that the universe is frozen [2, 11]. Is it a correct conclusion when we can not put a clock outside the universe?

\noindent         When we consider the universe as a whole, we encounter a conceptual paradigm shift. Time (using a clock outside the system), temperature (using a thermometer outside the system), entropy (the total number of ways to rearrange the internals of the system while keeping its external unaltered) and information (the degree of order) has been defined for a system which has boundary with the remained part of the universe. 

\noindent     When, we consider the universe as whole, a conceptual paradigm shift happen. Here, we should define new concepts for time, temperature, entropy, order (orderless), information and so on and pay enough attention to using the concepts which belong to a system and its environment. If we use these old concepts and laws which have satisfied for system and its environment and not the universe as a whole, we will find that the universe is frozen and we need a Big Bang or Big Bounce for describing the current measured experimental data [2, 11]. Of course in next section, we will explain how we need the Big Bang and or Big Bounce based on our definition of obligated minimum boundary between things and obligated minimum boundary between past and future.

\section{Ground Unified Reaction Platform}

\noindent Using the explanations in section II, we consider the below principles (axioms):

\begin{enumerate}
\item  The world is things which change i.e., it is variable things (events).

\item  Things interact with each other and their interactions and variations make the observables of the universe.

\item  A system is composed by variable things and boundaries.

\item  A boundary is the border of system which separates system from its environment (background).

\item  All things which are placed outside the boundary have negligible effects on the observables of the system while all things inside the boundary have mutual significant effects on each other.  

\item  A system composed by things is completely and conceptually different from the universe as a whole. 

\item  When we shift from a system to universe, a conceptual paradigm shift happen. During the usage of the concepts (such as time, temperature, entropy, information, {\dots}) and laws (second law of thermodynamic, relativity, quantum physics {\dots}) belong to a system when we deal with the universe as a whole, we should be very careful  and pay enough attention to the conceptual paradigm shift. It may be necessary; we redefine the concepts (e.g. time) and extract the laws (e.g. quantum gravity) again with compatibility with the definition of the universe.  

\item  There are two main interactions between things belong to a system. One is based on the biological conscious and other is based on quantum field theory including gravity or not. 

\item  The entangled interaction is a kind of conscious which can be called pre-conscious because the organism aware that a specific relation exist between other parts of the system before doing any experiment. Here, the different outcomes of the experiment can be determined with equal probability before doing the experiment (such as, half and half occurrence probability between entangled spin up and spin down electrons).  

\item  If the occurring probability (based on the quantum physics) is very high (such as classical physics limit) due to any reason (such as interactions between things of a system or interaction between system and its environment) and an organism is a part of the system, the conscious is the past-conscious since before doing the experiment one can guess the final results, deterministically (such as falling a ball from top of a tower). 
\end{enumerate}

\noindent     In continuing, we try to introduce a new concept which is called grand unified reaction platform (GURP) by considering the above axioms. Before doing that, we should review the concept of consciousness. The variable things of the universe can be categorized in two different branches which are organism and non-organism things. Organism things have the consciousness capability while the non-organism things have not. Here, the capability is composed by two concepts which are capable and ability. For example, a health two years old child has a hidden inherent learning aptitude  for learning the science but he/she should passes the different scientific courses for learning the  science.  It means that the emergence of each capability needs two main elements which are inherent aptitude and programing for maturing and improving the aptitude. Organism has the experience capability. Organism is experimenting every day through his/her five main senses which are seeing, hearing, touching, tasting, olfaction and emotion. It should be noted that [25, 26]: 

\noindent ``\textit{We usually explain what a thing does, how it changes and how it is put together, in science. But, an explanation of consciousness i,e., answering the question: why is it conscious (awake)?; goes beyond the method of science''}. 

\noindent Sometimes, our experience does not coincide with external reality. Let us to do an experiment by using two light emitting diodes (LED) in green and red colors. An observer who is not aware of these LEDs sees the flash of red LED for only 20 milliseconds (ms) followed immediately by the flash of green LED for 20 ms duration. What, he/she will report, is seeing a flash of yellow light [13]. It means that his/her central nervous system (SNC) integrates the two successive processes and in consequence he/she reports seeing the yellow color because there is a period of time during which he/she is aware of an event or an entity. It is called the duration of the present [13, 27]. In biology, time has physiological periodicities where future passes to past though the present (an infinitely thin boundary) [13].   

\noindent          For understanding and explaining the behavior of a system (not universe as whole) composed by non-organism variable things; an organism should interact with it. There are three cases as below:

\begin{enumerate}
\item  The organism is placed outside the boundary and its consciousness, which is emerged through its interaction with the system, coincides with the reality. Under the condition, he/she is able to explore the physical laws which govern over the system and inside its boundary. Here, his/her level of awareness (LoA) is complete. The observer is able to specify (guess) the output of his/her experiment deterministically (probabilistically). We prefer to call the LoA as past-awareness (pre-awareness and/or non-deterministic).  

\item  The organism is placed outside the boundary and its consciousness does not coincide with the reality. Since, he/she is not aware about his/her mistake the explored laws will be considered as govern physical laws over the system and inside its boundary. Here, his/he level of awareness (LoA) is not complete but the probability of his/her awareness about own mistake is not always zero and someday he/she will be aware and try to find the correct physical law. From the point of view, sometimes it seems that the physical laws have been changed by increasing his/her LoA. 

\item  Sometime the system and organism are not separable and in consequence the organism is placed inside the boundary of the system. Since, the explaining the consciousness goes beyond the method of science [25, 26], he/she is not able to explore the physical laws which govern over the system and inside its boundary. The situation is out of the scope of science. 
\end{enumerate}

\noindent        Now, let us to review the credibility border of current physical paradigms based on the above categorizations. By putting the organism outside the boundary of a system (not universe as a whole), we assume that the velocity of object is shown by $v$  and the velocity of light ($c$), the Planck constant ($\hbar$) and the gravity constant (G) stand for the special relativity effects, quantum physic effects, and gravity effects, respectively. Of course, the symbols only use for showing the  
paradigm shift. For both  $(v/c)$ and $\hbar\rightarrow 0$ , the Newtonian physics, for only $\hbar\rightarrow 0$ , the relativistic physics and  for only $(v/c)\rightarrow 0$, the quantum physics satisfies inside the boundary of the system. 
Also, for both  $(v/c)$ and $\hbar\nrightarrow 0$ , the quantum field theory satisfies inside the boundary. 
Of course in the above four cases, it is assumed that the gravity has no significant effect on the system (i.e., $G\rightarrow 0$ ) and in consequence it is placed outside their boundaries. Also in Newtonian and relativistic physics, the LoA is past-awareness and the observables follow the deterministic laws but in quantum physics and quantum field theory the LoA is pre-awareness and/or non-deterministic. In consequence, the observables do not follow the deterministic laws. It should be noted that, the pre-awareness about the observables causes the entanglement between them but does not remove their probabilistic behavior before doing the measurement, although by measuring one observable, the organism is able to specify its entangled observable with certainty i.e., deterministically (Fig. 2). When the system is not the universe of a whole, each physical law satisfies inside own boundary and credibility border. Therefore, in a part of the universe (not a universe as  a whole) by changing the credibility border of laws, the laws change due to the change of LoA.  Smolin has claimed that by considering the universe as a whole it seems that the laws have changed by passing the time[2].

\noindent 

\noindent 

\noindent 

\noindent 

\noindent 

\noindent 
\begin{figure*}[t!]
\includegraphics[width=\textwidth]{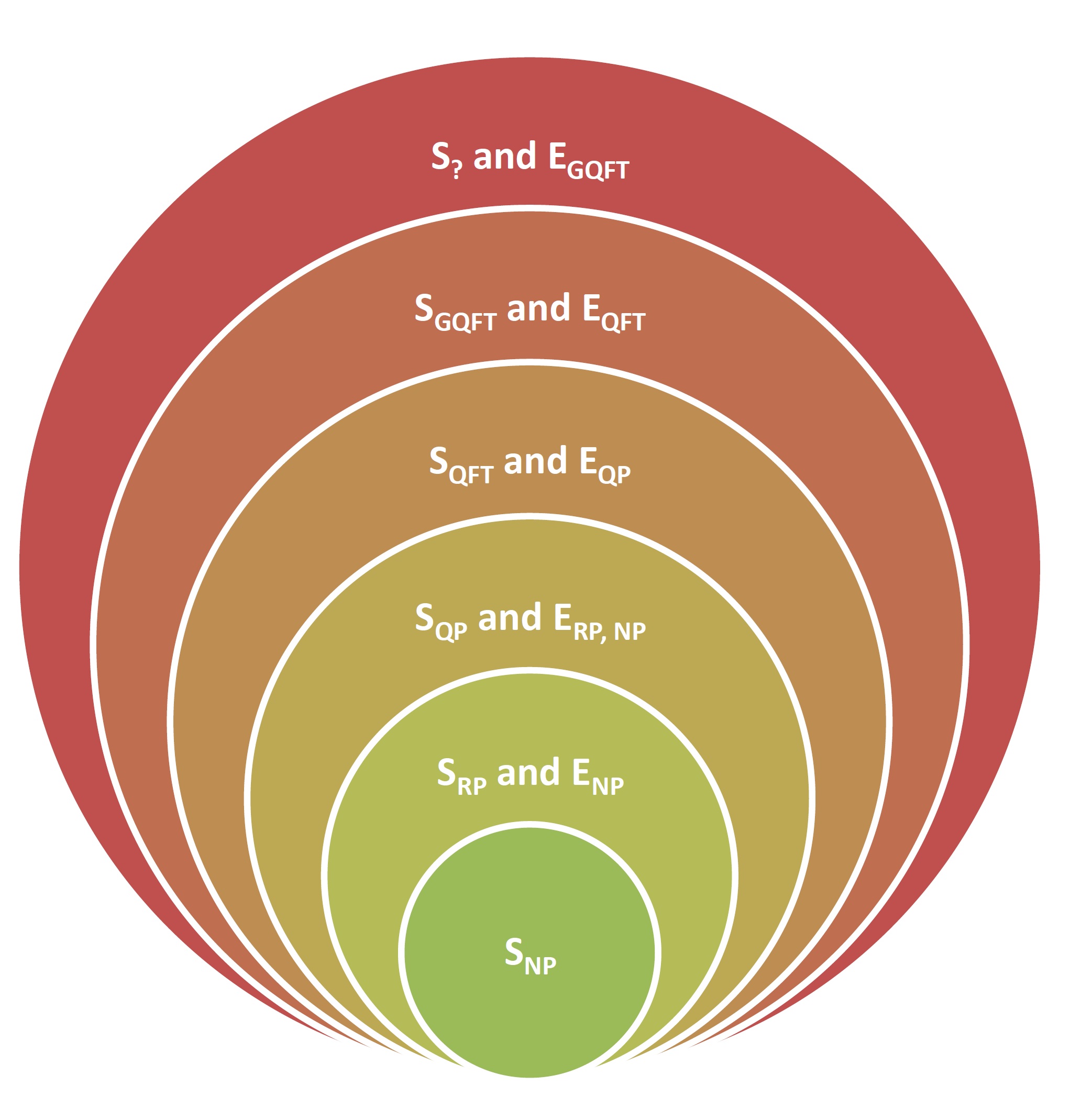}
\textbf{\caption{(Color online)The creditability border of different physical theories. Here,}${\boldsymbol{\ }\boldsymbol{S}}_{\boldsymbol{NP}}\boldsymbol{:}\boldsymbol{Both}\boldsymbol{\ }\boldsymbol{v/c}\boldsymbol{\ }\boldsymbol{\&\ }\boldsymbol{\hbar }\boldsymbol{\to }\boldsymbol{0}\boldsymbol{\ }\boldsymbol{and}\boldsymbol{\ }\boldsymbol{G}\boldsymbol{\to }\boldsymbol{0}\boldsymbol{\ }$\textbf{, }${\boldsymbol{S}}_{\boldsymbol{RP}}\boldsymbol{:}\boldsymbol{Both}\boldsymbol{\ }\boldsymbol{G}\boldsymbol{\ \&\ }\boldsymbol{\hbar}\boldsymbol{\to }\boldsymbol{0}\boldsymbol{\ }$\textbf{, }${\boldsymbol{S}}_{\boldsymbol{QP}}\boldsymbol{:}\boldsymbol{Both}\boldsymbol{\ }\boldsymbol{G}\boldsymbol{\ \&\ }\boldsymbol{v/c}\boldsymbol{\to }\boldsymbol{0}\boldsymbol{\ }$\textbf{, }${\boldsymbol{S}}_{\boldsymbol{QFT}}\boldsymbol{:}\boldsymbol{Both}\boldsymbol{\ }\boldsymbol{v/c}\boldsymbol{\ \&\ }\boldsymbol{\hbar}\boldsymbol{\nrightarrow }\boldsymbol{0}\boldsymbol{\ }\boldsymbol{and}\boldsymbol{\ }\boldsymbol{G}\boldsymbol{\to }\boldsymbol{0}$\textbf{, and }${\boldsymbol{S}}_{\boldsymbol{GQFT}}\boldsymbol{:}\boldsymbol{Both}\boldsymbol{\ }\boldsymbol{v/c}\boldsymbol{\ \&\ }\boldsymbol{\hbar}\boldsymbol{\nrightarrow }\boldsymbol{0}\boldsymbol{\ }\boldsymbol{and}\boldsymbol{\ }\boldsymbol{G}\boldsymbol{\nrightarrow }\boldsymbol{0}$\textbf{. Here, S and E stand foe system and environment, respectively. Also, NP, RP, QP, QFT, and GQFT are abbreviation of Newtonian physics, relativistic physics, quantum physics, quantum field theory, and gravitational quantum field theory, respectively.}}
\label{fig:Fig2}
\end{figure*}

\noindent But, what's about the universe as a whole? Before discussing about the subject, let us to review the biological immortality. Wikipedia says [28]: 

\noindent \textit{`` Biological immortality is a state in which the rate of mortality from senescence is stable or decreasing, thus decoupling it from chronological age. Various unicellular and multicellular species, including some vertebrates, achieve this state either through their existence or after living long enough. A biologically immortal living being can still die from means other than senescence, such as through injury, disease, or lack of available resources{\dots}.Biologists chose the word immortal to designate cell that are not subject to the Hayflick limit, the point at which cells can no longer divide due to DNA damage or shortened telomeres.''}

\noindent and also it says [29]:

\noindent \textit{`` Turritopsis nutricula is a small hydrozoan that once reaching adulthood, can transfer its cells back to childhood. {\dots}Hydrozoans have two distinct stages in their life, a polyp stage and a medusa stage{\dots}{\dots}..Generally in hydrozoa the medusa develops from the asexual budding of the polyp and the polyp results from sexual reproduction of medusa{\dots}{\dots} Turritopsis nutricula in any point of the medusa stage has the ability to transfer back into its polyp stage.''}

\noindent  Of course, some scientists [30, 31] have discussed about the Turritopsis nutricula and reversing the life cycle, respectively. 

\noindent        Now let us to assume that the universe is a Turritopsis nutricula (thing) plus a conscious observer (organism) whose presence has no significant effect on the thing.  What will be the observer's report about the changes in the thing? Depends on the observation conditions, he/she will report a cyclic change between polyp stage and medusa stage of the thing or will report that the thing is always juvenile. In the other word, when the measurement is always done at polyp stage i.e., at times, $=nT_0$ , where $T_0$ is the cyclic time and $n=1,2,3,\dots $, the observer see a frozen universe while when the measurement is always done at medusa stage i.e., at times, $(n-1)T_0<t<nT_0$, the observer see a universe that changes. When the measurement is done at infinitely thin boundary between medusa stage (future) and polyp stage (past) the observer see a cyclic change. The simple example has the below lessons for us:

\begin{enumerate}
\item  The remained things are separated from other things and conscious observer (organism) by a specific boundary which divides the universe to system (remained things) and its environment (other things and organism). The obligated smallest dimension of boundary (OSDB) between things under the observation conditions is called the ``quanta of space'' $\mathrm{\{}$Fig. 3(a)$\mathrm{\}}$.

\item  The changes of things are observable. If no change is observed the thing is frozen.

\item  The observed changes of things depend on the observation conditions.

\item  Based on the observation conditions, the changes of things can be classified in three categories called past, present, and future. 

\item  If an observation is always done at \textit{``polyp sate''} of the thing (past), the observer sees no changes and reports a frozen state of thing. But, if an observation is always done at \textit{``medusa state''} of thing (future), the observer sees changes and reports a variable (non- frozen) thing. 

\item  There is always an infinitely thin boundary between future and past which is called present. The duration of present (i.e., the width of the thin boundary) depends on the obligated smallest time duration (OSTD) under the observation conditions. The smallest time duration is called the ``quanta of time'' $\mathrm{\{}$Fig.3 (b)$\mathrm{\}}$.

\item  For better understanding the concept of the present stage we should pay much attention to a cyclic change. In a cyclic change, the future passes to the past though the present $\mathrm{\{}$Fig. 3(c)$\mathrm{\}}$.
\end{enumerate}

\noindent        

\noindent 

\noindent 

\noindent 

\begin{figure*}[t!]
\includegraphics[width=\textwidth]{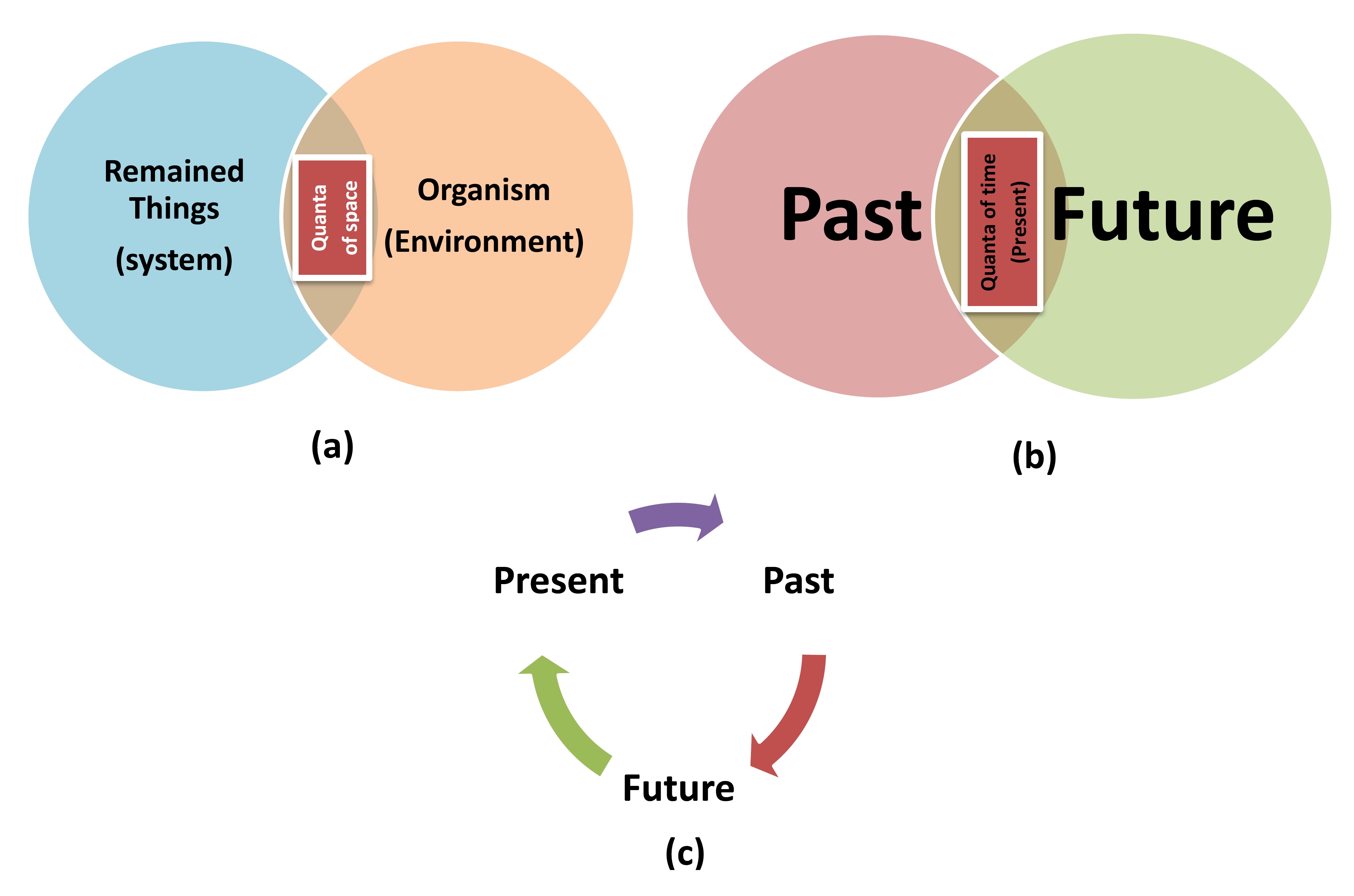}
\textbf{\caption{(Color online) (a) The quanta of space, (b) the quanta of time, and (c) a cyclic evolution}}
\label{fig:Fig3}
\end{figure*}

\noindent \textbf{   }   Now let us to mix the concepts of Fig2 with the concepts of Fig.3. In Newtonian universe since $(v/c)\to 0$, there is an obligated broad boundary between past and future. In consequence, the assumption of continues time (a stream of time) is meaningful. But, by increasing the ratio $(v/c)$ , the width of the obligated boundary becomes thinner.  When the ratio 
is equal or greater than a critical value, the width of the boundary approached to zero and past touches the future at a point which can be called the Dirac point. It is a singular point and in consequence we are not able to define an abstract time. After touching past with future the correct physical theory is relativistic theory.      However, not only in Newtonian physics but also in relativistic physics, we can recognize A-thing from B-thing with very high accuracy and without perturbing them and in consequence we can imagine that the universe contains things. It means that there is an obligated broad boundary not only between things but also between observer and observable in these theories. But by decreasing the width of the boundary, which can be shown mathematically as $\hbar\rightarrow 0$, the separation between them and also their non-perturbing interaction disappears gradually and we reach to the atomic scale. Here, the quantum physics is true theory which can be non-relativistic or relativistic. Therefore, the Planck constant  $\hbar$  can be proportional to the bench mark of the obligated smallest dimension of the boundary between things (minimum space).

\noindent     But, some questions can be asked. What are the nature of the obligated minimum boundary between things (minimum space) and the obligated minimum boundary between past and future (present)? Whether they are things if things are matter and fields? Whether the boundaries are made by the gravitational field? Based on the general relativity theory, the spacetime is gravitational field and the standard model, which is a non-gravitational quantum field theory, has unified mater with fields except gravitational field. Therefore, for all mentioned cases in the previous paragraph, the bench mark of gravitational field approaches to zero i.e., $G\rightarrow 0$ . But, if we accept (or assume) the boundaries are made by the gravitational filed what is about the case $G\nrightarrow 0$? At the beginning of the section, we assume that the universe is variable things as a whole, and in consequence it is not possible we consider other things (such as clock and, thermometer) outside it. It means that the state of things change and in consequence a set of data is generated. The content of the set depends on the existence of other things, the interaction between them, and their interaction with the gravitational field, if $G\nrightarrow 0$. Generally, based on the quantum physics, we first prepare the necessary conditions for doing a specific measurement and then measure the observables. Therefore, there are two different states which are state- prepared and state- measured [9, 10]. Now, if the system is not frozen and the observer (organism) has no significant on the system, each pair of events include state-prepared and state-measured.  The state of the pairs i.e., the space of data set is shown by $\mathfrak{H}$. It should be noted that, the space $\mathfrak{H}$ includes not only the data of observable variation but also the variations of gravitational field. In The other word, when $G\nrightarrow 0$ , we should not only measure the distance of the parts of the measurement apparatus and the time lapsed between them but also the variation of the physical observable [9, 10]. But for $G\nrightarrow 0$, it is not possible we separate the change of spacetime from the change of observables. In the other words, for $G\nrightarrow 0$ ($G\to 0$) we deal with one (two) measurement(s) because the mentioned two measurements are on the same ground [9, 10].

\noindent       Also based on the quantum physics, the transition probability amplitude can be shown as $W(s_{out},s_{in})$ where $s_{out}(s_{in})$ stands for state-measured (prepared) [9, 10].  However, if we can define a suitable inner scalar product then $W(s_{out},s_{in})$ will be equal to the inner scalar product between $s_{out}$ and $s_{in}$ i.e., their correlations ,$\ W\left(s_{out},s_{in}\right)=\left\langle s_{out}{\left|\vphantom{s_{out} s_{in}}\right.\kern-\nulldelimiterspace}s_{in}\right\rangle $. Of course, for calculating the dynamic of the system, we should find (define) the Hamiltonian as a function on ${H}$. For example, if operator $P$ stands for projection on the space including the solution of the WdW --equation, the transition amplitude will be $W\left(s_{out},s_{in}\right)=\left\langle s_{out}{\left|\vphantom{s_{out} P s_{in}}\right.\kern-\nulldelimiterspace}P{\left|\vphantom{s_{out} P s_{in}}\right.\kern-\nulldelimiterspace}s_{in}\right\rangle $ and the dynamic of the system is explained by a function on $\mathfrak{H}$ which is called Hamilton (not Hamiltonian) [9, 10]. Therefore, we can summarize the differences between physical paradigmsas what is shown in table 3.

\noindent 

\noindent 

\noindent 

\noindent

\begin{table}[h]
\caption {\textbf{The differences between physical paradigms. GQFT, QFT, RQP, QP, RM, and NM are abbreviations of gravitational quantum field theory, quantum field theory, relativistic quantum physics, quantum physics, relativistic mechanics, and Newtonian mechanics, respectively.}} 
\centering
\begin{tabular}{|p{0.9in}|p{1.0in}|p{1.2in}|p{0.9in}|p{1.0in}|p{0.9in}|} \hline 
 \textbf{Paradigm} & \textbf{Data space (}$\mathfrak{H}$\textbf{)} & \textbf{Transition amplitude} & \textbf{Hamiltonian} & \textbf{OSDB} & \textbf{OSTD} \\ \hline 
\textbf{GQFT} & Collection of both state-prepared and state-measured & $W\left(s_{out},s_{in}\right)=\left\langle s_{out}\mathrel{\left|\vphantom{s_{out} s_{in}}\right.\kern-\nulldelimiterspace}s_{in}\right\rangle $ & Hamilton function is defined on $\mathfrak{H}$ [9, 10] & There is a minimum gap between things & There is a minimum gap between past and future \\ \hline 
\textbf{QFT (RQP)} & Collection of state-measured at specific points of spacetime & $W\left(s_{out},s_{in}\right)=\left\langle s_{out}\mathrel{\left|\vphantom{s_{out} P s_{in}}\right.\kern-\nulldelimiterspace}P\mathrel{\left|\vphantom{s_{out} P s_{in}}\right.\kern-\nulldelimiterspace}s_{in}\right\rangle $\newline $P$ stands for projection on the space including the solution of the special equation [9, 11]\newline  & Hamiltonian function is defined on $\mathfrak{H}$ & Gravitational field has no significant effect & Future touches the past at Dirac point \\ \hline 
\textbf{QP} & Collection of state-measured at specific points for specific times & Same as QFT & Same as QFT & A continuous space is considered. & A stream of time is considered \\ \hline 
\textbf{RM} & Same as QFT & Things change deterministically & Can be defined [9, 10] & A continuous spacetime is considered & Future touches the past at Dirac point \\ \hline 
\textbf{NM} & Same as QP & Things change deterministically & Can be defined [9, 10] & The world contain things & A stream of time is considered \\ \hline 
\end{tabular}
\end{table}
\noindent 

\noindent 

\noindent 

\noindent 

      Based on our best current knowledge, the universe is things which are matter, fields of standard model plus gravitational field.  Some important subjects about our current knowledge are:

\begin{enumerate}
\item  We measure the observables of the universe by using the equipment and tools which have been made by using the current physical theories and they work based on them. The theories are satisfied over a part of the universe and on the specific conditions. In the other words, they satisfy over a region of universe which is separated from the remained part of the universe by some boundaries.  

\item  Since the gravitational field is a part of the universe, the data set include state-prepared and state-measured. Therefore, there is not only an obligated minimums boundary between things but also an obligated minimum boundary between past and future (future does not touch past at Dirac point) when the effect of the gravitational field is significant on the boundaries.

\item  We are a part of the universe and it seems that we stand on the future when we measure the cosmological radiations which belong to our past, i.e., they have traveled a long distance in the cosmos. Therefore, the universe as a whole is at its ``\textit{medusa} stage'' now and we see the universe as variable things. It should have a beginning point (Big Bang). As we see, the govern laws on Big Bang conditions differ from govern laws on sometimes after Big Bang( such as our current conditions) since things, the interactions between them and with gravitational filed had changed. The interaction between things and between things and gravitational field create new things (mater and or fields) in the world and in consequence it is expected that the govern laws on the universe change by changing the universe (set of data i.e. state-prepared plus state-measured). Then it seems that a flow of time exist in the universe which separates the future from the fast. Under our current situation we are trying to repeat the Big Bang condition (e.g., Long Hadronic Collider (LHC) set up) and find the suitable gravitational quantum field theories.   

\item  If universe as a whole is at its ``polyp stage'' it means that it is always in past i.e., it is frozen. The frozen universe may be explained by a correct developed version of WdW-equation probably. 

\item  If universe has a cyclic evolution, the future passes through the past and in consequence we should have a Big Bounce instead of Big Bang. If it is true there is thin boundary between past and future and the experimentalists should find some cyclic phenomena and their rhythms at the scale of whole universe.
\end{enumerate}

\noindent Anyhow, it seems that a reaction platform can be made by using the below materials:

\begin{enumerate}
\item  Events (variable and interacting things) which are mater, fields of standard model and gravitational field

\item  A boundary separates system (things inside the boundary) from its environment (means things outside the boundary). The outside things have no significant effect on the inside things.  

\item  An obligated boundary between things exists and its minimum can be effected by the gravitational field (specially at Planck scale).

\item  An obligated boundary between past and future exists and its minimum can be effected by the gravitational field (specially at Planck scale).

\item  A conscious organism (observable) which has no significant effect on data set (state-prepared plus state-measured) is placed outside the system.

\item  The observables are variation of things and their interactions.

\item  If a conscious organism is placed inside the boundary the science cannot explain the system due to its inability to explain the consciousness.   
\end{enumerate}

\noindent 

\noindent How the above materials are used and what are their preconditions will specify the type of physical theory under the used condition. When the gravitational filed has significant effect the reaction platform is called Grand Unified Reaction Platform (GURP). Fig. 4 shows the GURP and its different branches.  

\noindent The grand unified reaction platform suggest us to think about a new theory based on the above mentioned principles which can categorize different physical paradigms based on their differences in the obligated minimum boundary between things, between past and future and between things and conscious organism. We prefer to call the new theory as  "Boundaries Theory". In boundaries theory we should show how the different physical paradigms can be extracted, the gravitational filed acts on the boundaries and how the different boundaries can be transferred to each other using a suitable transformation.      
\begin{figure*}[t!]
\includegraphics[width=\textwidth]{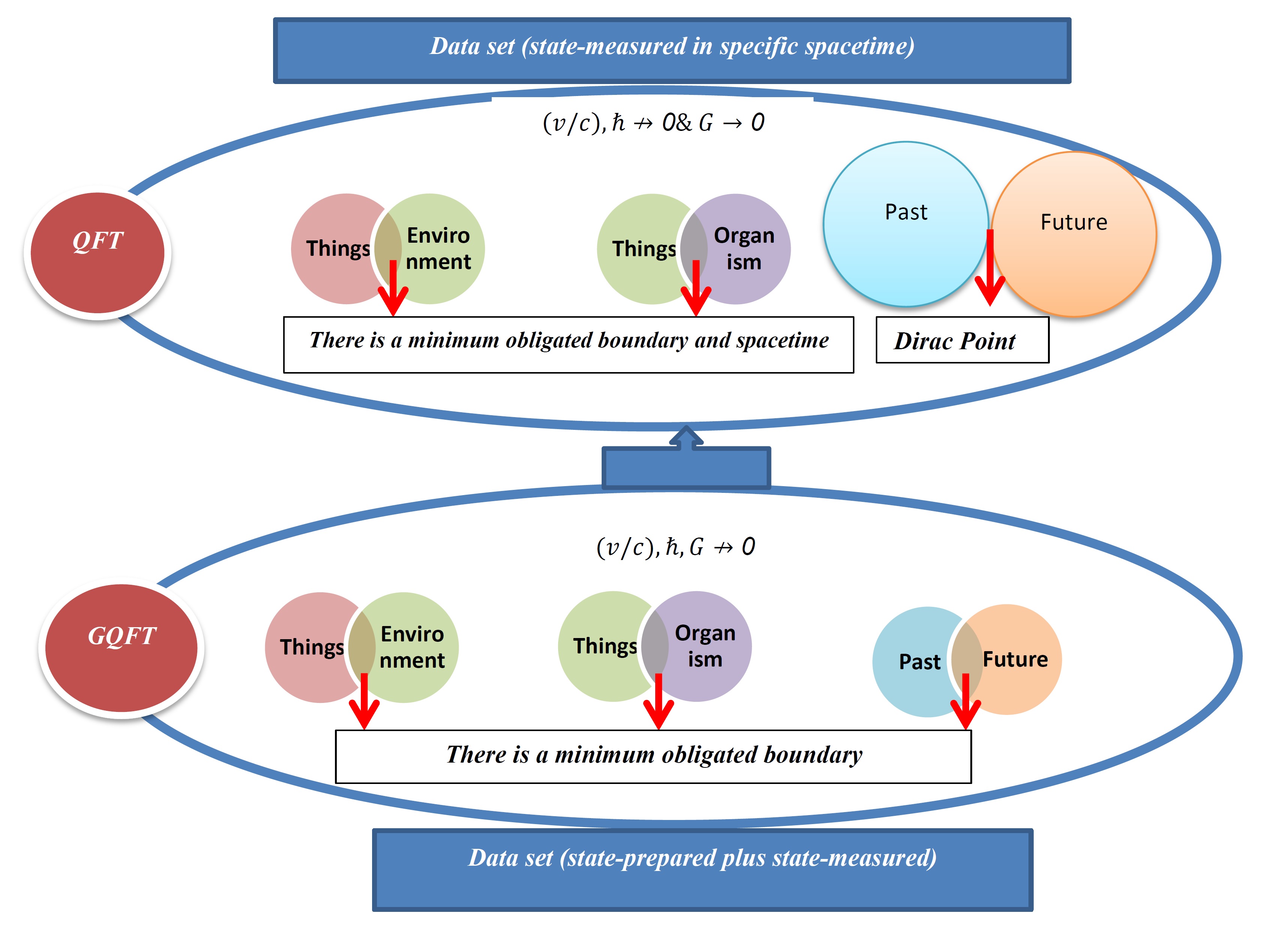}
\textbf{\caption{Fig.4 (Color online) The Grand Unified Reaction Platform for building a physical theory for universe}}
\label{fig:Fig4}
\end{figure*}

\section{Summary}

\noindent          Universe is variable things. The observables are variation of things and their interactions. Organism is a conscious thing.  Science can explain conscious but cannot explain the consciousness. A boundary divides a system to two parts called things and environment. The things which are placed outside the boundary has no significant effect on the inside things. If organism is placed inside the boundary the science cannot explain the system due to its inability to explaining the consciousness.  The boundary between things has an obligated minimum value which is called minimum space. The obligated minimum boundary between past and future is called present. The gravitational field has significant effect on the present and the obligated minimum boundary between things in Planck scale. They are the necessary material for manufacturing a reaction platform for explaining the current situation of different physical paradigms respect to each other. We hope that the article can motivate the mathematical physicists to work on the subject and provide the necessary infrastructure for next physical followers .       

\noindent

\noindent 

\noindent 

\noindent 

\noindent 

\noindent 

\noindent 

\noindent 

\noindent 

\noindent 
\noindent 

\noindent 

\noindent 

\noindent 

\noindent 

\noindent 

\noindent 

\noindent 

\noindent 

\noindent 
\clearpage
 \textbf{References}

\noindent [1] Julian Barbour, ``The End of Time'' (Oxford university press, 1999)

\noindent [2] Lee Smolin, ``Time Reborn'' (Houghton Mifflin Harcourt publishing company, 2013)

\noindent [3] Paula Marchesini, ``The end of Time or Time Reborn? Henri Bergson and the Metaphysics of Time in Contemporary Cosmology'', Philosophy and Cosmology, Vol.21 (2018)

\noindent  [4] Henri Bergson, ``The creative Mind: An introduction to metaphysics'' (Trans. Mabelle L. Andison, New York: Citadel, 1964)

\noindent [5] Henri Bergson, ``Creative Evolution'' (Trans. Arthur Mitchell. Solis Press. Kindle Edition, 2014)

\noindent [6] Carlo Rovelli, ``The strange equation of quantum gravity'', Class. Quantum Grav. 32, 124005 (2015)

\noindent [7] Lewis H. Ryder, ``Quantum Field Theory'' (Cambridge University Press, 1996)

\noindent [8] Robert F. Pierret, ``Advanced Semiconductor Fundamentals'' (Prentice Hall, 1967)

\noindent [9] Carlo Rovelli, ``Quantum Gravity'' (Cambridge University Press, 2003)

\noindent [10] Carlo Rovelli and Francesca Vidotto, ``Covariant Loop Quantum Gravity'' (2013)

\noindent [11] Carlo Rovelli, ``The order of time'' ( Cambridge University Press, 2015)

\noindent [12] Joseph Conlon,'' Why String Theory?''  (CRC Press, 2017)

\noindent [13]B. Gunther and E. Morgado, `` Time in Physics and   biology'', Biol. Res. 37, 759 (2004).

\noindent [14] J. S. Huxley, ``Problems of Relative Growth'' (Methuen, London, 1932).

\noindent [15] B. Gunther and E. Morgado, ``Body mass and body weight: a dual reference system in biology'', Rev. Chil.  Hist.  Nat. 76, 57 (2003).

\noindent [16]  Deborah Netburn Los Angeles Time, 24, July, 2014 and Pranab Roy and Prittam Goswami, `` Freeze tolerance in wood frogs'', J. Invest. Genomics, 6, 1 (2019)

\noindent [17] S. Bulterijs, R. S. Hull, V. C. E. Bjork, and A. G. Roy, `` It is time to classify biological ageing as a diseases'' Front. In Genet. 6, 205 (2015).

\noindent [18] D. Callahan and E. Topinkova, ``Is aging a preventable or curable disease?'', Drugs Aging, 13, 93 (1998).

\noindent [19] L. Hayflick, ``Biological aging is no longer an unsolved problem'', Ann. N. Y. Acad. Sci., 1100, 1 (2007).

\noindent [20] C. Lopez-Otin, M. A. Blasco, L. Paartridge, M. Serrano, and G. Kroemer, ``The hallmarks of aging'', 

\noindent  Cell, 153, 1194 (2013)

\noindent [21] ScienceDaily, ``Disrupting genetic process reverses aging in human cell'' (Sept.13/2018)

\noindent [22] MIT Technology Review, Biotechnology, ``Scientists hope to lengthen dog years'' (Oct.22/2015).

\noindent [23] M. Ullah and Z. Sun,''Stem cells and anti-aging genes: double-edged sword-do the same job of life extension'',  Cell Reasr. Thera., 9, 3 (2018).

\noindent [24] J. Waterhouse, J. Royal Society of Medicine, 8, 398 (1992)

\noindent [25] Stephen Grossberg,'' Toward solving the hard problem of consciousness: The varieties of brain resonance and the conscious experience that they support'',  Neural Networks, 87, 38 (2017)

\noindent [26] L. N. Long and T. D. Kelley,''Review of Consciousness and the Possibility os Conscious Robots'',  J. Aerospace Comput. infor. Commu., 7, 68 (2010)

\noindent  [27] R. Efron, Ann. NY Acad. Sci., 38, 713 (1967)

\noindent [28] http://en.wikipedia.org/w/index.php?title=Biological\_immortality\&oldid=910723360

\noindent [29] http://en.wikipedia.org/w/index.php?title=Turritopsis\_nutricula\&oldid=899540367

\noindent [30] G. Bavestrello, C. Sommer, and M. Sara, ``Bi-directional conversion in Turritopsis nutricula (Hydrozoa)'', Science 56, 137 (1992)

\noindent [31] S. Piraino, F. Boero, B. Aeschbach, and V. Schmid, `` Reversing the Life Cycle: Medusa Transforming into Polyps and Cell Transdifferentiatin in Turritopsis nutricula'',  Biol. Bull. 190, 302 (1996)

\noindent

\end{document}